\documentclass[pre,aps,onecolumn,floatfix,superscriptaddress,10pt]{revtex4-1}

\usepackage{amsmath}
\usepackage{amsfonts}
\usepackage{amssymb}
\usepackage{graphicx}
\usepackage{esvect}
\usepackage[hyperindex,colorlinks,hyperfootnotes = false,citecolor = black,]{hyperref}
\usepackage[usenames]{color}

\renewcommand{\vv}[1]{\overrightarrow{ #1}}

\begin{document}

\title{Using higher-order Markov models to reveal flow-based communities in networks}

\author{Vsevolod Salnikov}
\email[E-mail: ]{vsevolod.salnikov@unamur.be}
\affiliation{naXys, University of Namur, Rempart de la Vierge 8, 5000 Namur, Belgium}
\author{Michael T. Schaub}
\email[E-mail: ]{michael.schaub@unamur.be}
\affiliation{naXys, University of Namur, Rempart de la Vierge 8, 5000 Namur, Belgium}
\affiliation{ICTEAM, Universit\'e catholique de Louvain, Avenue George Lema\^itre 4, B-1348 Louvain-la-Neuve, Belgium}
\author{Renaud Lambiotte}\email[E-mail: ]{renaud.lambiotte@unamur.be}
\affiliation{naXys, University of Namur, Rempart de la Vierge 8, 5000 Namur, Belgium}

\begin{abstract}
Complex systems made of interacting elements are commonly abstracted as networks, in which nodes are associated with dynamic state variables, whose evolution is driven by interactions mediated by the edges. 
Markov processes have been the prevailing paradigm to model such a network-based dynamics, for instance in the form of random walks or other types of diffusions.
Despite the success of this modelling perspective for numerous applications, it represents an over-simplification of several real-world systems. 
Importantly, simple Markov models lack memory in their dynamics, an assumption often not realistic in practice.
Here, we explore possibilities to enrich the system description by means of second-order Markov models, exploiting empirical pathway information.
We focus on the problem of community detection and show that standard network algorithms can be generalized in order to extract novel temporal information about the system under investigation. 
We also apply our methodology to temporal networks, where we can uncover communities shaped by the temporal correlations in the system.
Finally, we discuss relations of the framework of second order Markov processes and the recently proposed formalism of using non-backtracking matrices for community detection.
\end{abstract}

\maketitle

\section*{Introduction}
Dynamics on complex networks, such as the diffusion of information in a social networks, are commonly modelled as Markov processes.
An advantage of this approach is that for every (static) network with positive edge-weights we can define a corresponding Markov process by interpreting the network as the state space of a random walker, and assigning the state-transition probabilities according to the link weights.
This direct correspondence between the state space of the Markov process and the network enables us to examine the interplay between structure and dynamics from two sides.
On the one hand, one can assess how the topological properties of a network influence the dynamical process.
On the other hand, this coupling between topology and dynamics allows us to explore the structure of a network by means of a dynamical process. 
Specifically, for a linear Markov process the impact of the network structure on the dynamics will be mediated by the spectral properties of the matrix governing the time-evolution of the process, e.g. the adjacency matrix or the Laplacian \cite{Newman2010, Lambiotte2015}.
Reversely, spectral properties can be used to uncover salient structural properties of a network, such as modular organisation \cite{Delvenne2010,Schaub2012,Newman2013}.

While simple Markov models have been very successful in modelling dynamics of complex systems and found many applications, they have one obvious disadvantage.
In this class of models, the future state of the system only depends on its current state and does not account for its history. 
In a diffusion process, for instance, the next position of a random walker only depends on the currently occupied node and its outgoing links, but not on any of the previously visited nodes. 
However, as it has been emphasised recently, for a broad range of networked systems, flows tend to exhibit a temporal path dependence \cite{Rosvall2014,Scholtes2014}. 
Think of human mobility: the places a person is likely to visit next, often depend strongly on where the person came from.
For instance, a person coming to work from home is likely to return home afterwards \cite{Song2010}.
Other examples of processes with temporal memory include web traffic, journal citation flows and email cascades.
Such processes therefore cannot be reproduced accurately by simple Markov models.
However, the impact of this temporal correlations can often be well-approximated already by second-order Markov ($\mathcal{M}_2$) models \cite{Rosvall2014}.
Importantly, the transition probabilities to define these models can be obtained empirically by measuring pathways of interaction cascades, rendering such an approach suitable for applications like information spreading or human mobility.

\begin{figure}
\includegraphics{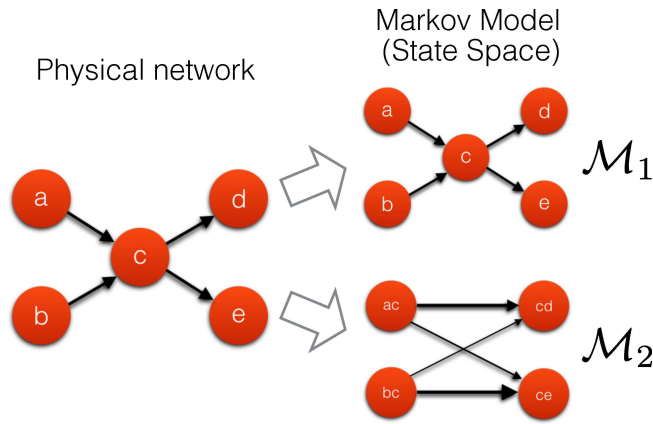}
\caption{\textbf{Schematic -- dynamics on networks as Markov processes}. A process on a physical network (left) may be abstracted in a number of different ways -- often this is done via a Markov model. In a first order Markov model $\mathcal{M}_1$ (right side, top panel), the state space is isomorphic to the physical network: every node corresponds to one state, every link indicates a transition between those states. In a second order Markov model the state space is structured like a \textit{directed line-graph} of the original network. The states in this $\mathcal M_2$ network can be identified with the directed edges in the original network and are connected if it is possible to traverse from one edge to another edge in  the original directed network. Note that, when projecting back these $\mathcal M_2$ dynamics onto the physical network, the probability to move from one node to another will thus appear non-Markovian.}
\label{fig1}
\end{figure}

In the standard Markovian network model ($\mathcal{M}_1$), the elementary states are identified with the nodes of the original network (see Figure \ref{fig1}).
In the $\mathcal{M}_2$ model, elementary states correspond to sequences of two nodes, and thus can be identified with \textit{directed edges} in the original network (see Figure \ref{fig1}).
Therefore we may think of a $\mathcal M_2$ model alternatively as a random walk between directed edges of the original network.
The state space of the $\mathcal{M}_2$ model defines a new network describing the observed dynamics, which we call here the $\mathcal{M}_2$ or memory network. 
The structural properties of the memory network can now be studied by many tools of network science, allowing us to uncover interesting patterns of flow in an $\mathcal{M}_2$ model.  

The purpose of this paper is to highlight how the formalism of higher order Markov models provides a simple means to extend network theoretic tools to take into account important dynamical properties as encoded in the $\mathcal{M}_2$ representation.
Indeed many network theoretic tools may be understood as the outcome of a matrix iteration or eigenvector computation \cite{Estrada2010}, which can be naturally associated to a dynamical process.

In this paper we focus on the problem of community detection, though higher-order models can be equally applied to other problems, including the assessment of node centralities \cite{Estrada2010,Rosvall2014} or the speedup (or slowdown) of spreading processes \cite{Scholtes2014}. 
While Ref. \cite{Rosvall2008} discussed how the map equation can account for higher-order flows in community detection, here our main example will be the Markov stability \cite{Delvenne2010,Delvenne2013,Lambiotte2014} formalism, as it incorporates many commonly used community detection measures as special cases, notably, the concept of Modularity \cite{Newman2006}, diverse Potts models \cite{Reichardt2004}, and spectral clustering \cite{Fiedler1975,Shi2000}.
We thereby highlight that all these classical algorithms can indeed be naturally generalised to account for memory effects by using $\mathcal M_2$ networks, showing that these representations provide a general tool for the analysis of a dynamics occuring on a latent network structure. 
Moreover, as spectral clustering can be shown to be a special case of the Markov Stability formalism for long times, we can draw further connections to recently proposed techniques of graph partitioning based on non-backtracking random walks \cite{Krzakala2013}, and discuss how these can be understood from the point of view of second order Markov processes.
We remark that as clustering an $\mathcal{M}_2$ network is equivalent to finding a partition of the edges of the original system, it naturally leads to the detection of overlapping communities in the same way as the analysis of line graphs~\cite{Evans2009}.
This can be a highly desirable feature, especially when analysing social networks, which tend to be organised in overlapping social circles \cite{Friggeri2011,Ahn2010}.

The remainder of this article is organised as follows. 
Initially, we review the Markov stability formalism for community detection, a general framework to detect flow-based communities in complex networks, and particularly emphasise its properties in the case of directed networks \cite{Lambiotte2015}. 
We show how this quality function can be naturally generalised for the analysis of $\mathcal{M}_2$ networks, which we illustrate by studying a flight network of the United States from the perspective of second order Markov models.
In this context, we also discuss the possibility to generate realistic pathway data using models of second-order Markov processes, even if only time aggregated network information is available.
Subsequently, we propose a mechanism to extract pathway statistics to build second-order models from event-based, temporal network data, which do not contain pathway statistics \textit{a priori}.
Using computer-generated data and time-resolved interactions records of school-children, we demonstrate how we can uncover communities which capture important flow-constraints imposed by the temporal activation patterns of the links.
Finally, we draw connections between the $\mathcal M_2$ network representation and non-backtracking random walks to uncover communities in sparse networks.

\section*{Markov Dynamics and Community detection}

\subsection*{Random walks on directed networks}
Let us consider a continuous-time random walk on a network with $N$ nodes, governed by the following Kolmogorov forward equation:
\begin{equation}
\label{ctrw}
\dot{p}_{j}(t) = - p_j(t) + \sum_{i} p_{i}(t) \frac{A_{ij}}{k_i^{\rm out}}.
\end{equation}
Here $p_{j}(t)$ denotes the probability of a walker to be present on node $j$ at time $t$; $A_{ij}$ is the weight of the link from node $i$ to node $j$, i.e., $A$ is the (weighted) adjacency matrix; finally, $k_j^{\rm out}$ is the weighted out-degree of node $j$.

Clearly this dynamics is driven by the normalised Laplacian matrix $L$, defined as $L_{ij} = - \delta_{ij} + A_{ij}/k_i^{\rm out}$.
The Perron-Froebenius theorem guarantees that a unique stationary solution exist for the above dynamics if the network is strongly connected.
For an undirected network, this stationary solution is $\pi_j=k_j/2M$, where $k_j=k_j^{\rm out} = k_j^{\rm in}$ is the node degree and $M$ is the total weight of the undirected links. 
For a general directed network the stationary solution is given by the dominant (left) eigenvector of $L$, which depends on the global network properties and cannot be expressed by a simple analytical formula. 

\paragraph*{Making the dynamics ergodic (when it is not).}
\label{ergodic} 

Note that for the continuous-time random walk defined above, a strongly connected graph implies an unique stationary distribution and an ergodic dynamics.
These criteria are a common requirement for dynamics-based methods for network analysis \cite{Delvenne2010,Rosvall2008}.
However, in a majority of real-world systems the system contains not just one but several strongly connected components (SCCs), whose sizes typically have different orders of magnitude.
A standard solution is to address this issue by either neglecting some nodes and focusing only on the largest SCC only, or by making the system strongly connected through random teleportations \cite{Lambiotte2012}. 
The first solution has the advantage of not distorting the dynamics, while the second allows for the analysis of the whole system. 
When analyzing real-world systems in the following, we will for these reasons use the first solution if the vast majority of nodes belongs to the largest SCC, while we will opt for the second solution if the system comprises a large number of disconnected components. 

\subsection*{Community structure from a dynamical viewpoint}
The structure of a network has a strong effect on the dynamics of a diffusing random walker: in unstructured (random) networks, the diffusion process will evolve almost isotropically and the walker will quickly reach its stationary distribution.
However, if the network contains structure, a walker can get trapped inside a group of nodes for a time far longer than expected. 
Such groups of nodes thus constitute flow-retaining, \textit{dynamical communities} in the network.
Hence a random walk dynamics can effectively be used to define a quality function for a network partition based on the persistence of the diffusion inside the groups.
By optimising such a quality function, we can therefore search for a modular partition of the network.
Important examples for such an approach include the map equation \cite{Rosvall2008} and the Markov stability framework \cite{Delvenne2010}, which we consider in this paper.
Note that, as highlighted in Refs. \cite{Rosvall2008,Schaub2012} this notion of community is markedly different to the commonly considered structural viewpoint of communities, in that we are interested in flow-retaining, rather than densely (homogeneously) connected substructures.

The Markov stability of a partition $\mathcal{P}$ is defined as:
\begin{equation}
R(t,\mathcal P) =\sum_{C \in \mathcal{P}}  P(C,t) - P(C,\infty)
\label{eq:modDeff}
\end{equation}
where $P(C,t)$ is the probability for a walker to be in community $C$ initially and at time $t$, for a system at stationarity.
Note that as all information about the initial position is lost after infinite time, $P(C,\infty)$ also describes the probability of two independent walkers to be in $C$.

For the process \eqref{ctrw}, the Markov stability quality function can be written explicitly as:
\begin{equation}
\label{Rt}
R(t,\mathcal P) =   \sum_{C \in \mathcal P} \sum_{i,j \in C} \left[ \pi_i \left( e^{t L} \right)_{ij}  - \pi_i \pi_j \right],
\end{equation}
where $e^{t L} $ denotes the matrix exponential.
Intuitively, with increasing time the walker will be able to explore larger and larger parts of the graph. 
The larger the time, the larger the modules that will be found in general.
By `zooming' through different Markov times one can thus reveal adequate scales of robust community structure, which manifest themselves as plateaux in time where the same partitions are found consistently \cite{Schaub2012,Delmotte2011}.
In the limit $t \rightarrow \infty$, it can be shown by eigenvalue decomposition that $R(t,\mathcal P)$ is  maximised by a bi-partition in accordance with the normalised Fiedler eigenvector, a classic method in spectral clustering~\cite{Shi2000}. 
Interestingly, the expression for Markov stability can be rewritten as the Newman-Girvan modularity \cite{Newman2006} of a \textit{different} network, whose adjacency matrix is given by
\begin{equation}
Y_{ij}=\dfrac{1}{2}\left[\pi_i \left( e^{t L} \right)_{ij} +\pi_j \left( e^{t L} \right)_{ji}\right].
\end{equation}
Here $\pi_i \left( e^{t L} \right)_{ij}$ is the flow of probability from $i$ to $j$ after a time interval $t$ at stationarity. 
In this new network $Y$, the weight of a connection between two nodes is thus modulated according to its importance for the diffusion dynamics.  
This rewriting of Markov stability as the modularity of a different network opens the possibility to use any modularity based algorithm for the optimization of stability, too.
For the results in the present manuscript, we have made use of the Louvain algorithm \cite{Blondel2008}, a fast greedy optimisation heuristic that has been shown to have a good performance in the literature.

In practice, it can be more efficient to consider (\ref{Rt}) in the limit of small times $t \rightarrow 0$. 
Performing a Taylor expansion and keeping only linear terms in $t$, results in the linearised Markov stability:
\begin{equation}
\label{Qt}
Q(t) =   (1-t) +  \sum_C \sum_{i,j \in C} \biggl[ t  \frac{\pi_iA_{ij}}{k_i^{\rm out}}   - \pi_i \pi_j \biggr].
\end{equation}
Note that for an \textit{undirected} network, we have $\pi_i=k_i/2M$ and $k_i^{\rm out}=k_i$, and one recovers the modularity of the original network for $t=1$ as well as the Potts-model heuristic \cite{Reichardt2004} for other Markov times (see \cite{Delvenne2010,Delvenne2013} for details).
However, for \textit{directed} networks \eqref{Qt} is not equivalent to the directed modularity of the original network, but is given by:
\begin{equation}
Y_{ij} = \dfrac{1}{2}\left[\pi_i \frac{A_{ij}}{k_i^{\rm out}} +\pi_j \frac{A_{ji}}{k_j^{\rm out}}\right].
\end{equation}
Interestingly, similar adjusted adjacency (or Laplacian) matrices have been proposed as means for community detection in the literature based on different reasoning~\cite{Chung2005,Satuluri2011}.

Note that, as with all applications of unsupervised algorithms, care is needed when interpreting the results. 
For instance, the optimisation of Markov Stability for a fixed Markov time inherits the limitations of modularity maximisation, such as a resolution limit \cite{Fortunato2006}, and a high degeneracy of optimial solutions  \cite{Good2010}. 
Nevertheless, this class of method has been applied successfully in numerous applications \cite{Fortunato2010}. 
Indeed, some of these problems can be circumvented by carefully sweeping through different resolution parameters \cite{Schaub2012,Schaub2012b} and finding partitions that are robust over a range of parameters \cite{Schaub2012,Delmotte2011,Lambiotte2014,Lambiotte2010}.
Most importantly, however, the notion of community inherent to the methods presented here is based on flows \cite{Delvenne2010,Schaub2012,Rosvall2008}.
This is in contrast to methods like stochastic blockmodels (SBM), see e.g. Ref. \cite{Holland1983,Karrer2011,Peixoto2014}, which employ a generative model for the whole network and aim at finding patterns of pairwise connections in the system, though very recent work aims to incorporate dynamical aspects within the SBM framework~\cite{Peixoto2015}.

\section*{Dynamical community detection using higher-order Markov models}
\subsection*{From first to second order Markov models}
Let us now show how we can incorporate second-order Markov models into the Markov stability framework.
As higher-order Markov models provide more faithful representations of the dynamics observed in real world systems, this enables us to capture more of the real flow constraints in the uncovered communities.  
For simplicity let us initially consider undirected networks composed of $N$ nodes and $M$ links.
The dynamics of a second order Markov process are encoded by the transition matrix $T(\vv{ij} \rightarrow \vv{jk})$,
which describes the probability that a walker moves from node $j$ to node $k$ if it came from $i$ in the previous step.
By definition, this transition matrix is normalised such that $\sum_k T(\vv{ij} \rightarrow \vv{jk}) = 1$. 

As sequences of two vertices in our original network correspond to the nodes in the $\mathcal M_2$ network, the entries in $T$ describe precisely the transitions from $\mathcal{M}_2$-node $\vv{ij}$ to $\mathcal{M}_2$-node $\vv{jk}$.
The (second-order) $\mathcal M_2$ process on the original network is thus equivalent to a first-order Markov process, albeit on a different network: namely, the $\mathcal{M}_2$ network composed of $2M$ nodes.
As each undirected link of the original network can be traversed in two distinct directions (from $i$ to $j$, and vice versa) it accounts for 2 nodes in the $\mathcal{M}_2$ network, $\vv{ij}$ and $\vv{ji}$. 
This implies that the $\mathcal M_2$ network is \textit{directed} even if the original network is undirected.
To see this, observe that if $k\neq i$, there cannot be a link between $\vv{jk}$ and $\vv{ij}$, even if a transition between $\vv{ij}$ and $\vv{jk}$ exists. 
Henceforth we use Greek letters to denote $\mathcal{M}_2$ nodes, and Latin characters to denote the nodes in the original $\mathcal{M}_1$ network.

Similarly to (\ref{ctrw}), we can define a continuous-time random walk on the $\mathcal{M}_2$ network as:
\begin{equation}
\dot{p}_\beta(t) = - p_{\beta}(t) + \sum_{\alpha} p_\alpha(t) T_{\alpha \beta} ,
\end{equation}
where $p_\alpha(t)$ is the probability of finding a walker on $\mathcal{M}_2$ node $\alpha$ at time $t$. 
Likewise the stationary distribution $\pi_\alpha$ of this process is given by the left eigenvector of the corresponding Laplacian, associated with eigenvalue $0$:
\begin{align}
\sum _\alpha \pi_{\alpha} [T_{\alpha\beta} - \delta_{\alpha\beta}] =0.
\end{align}

Note that we can also observe how a simple $\mathcal M_1$ Markov dynamics evolves from the point of view of an $\mathcal M_2$ network, i.e., we can view the $\mathcal M_1$ dynamics from the point of view of transitions between the (directed) edges of the graph.
As this is equivalent to lifting the $\mathcal M_1$ dynamics into the larger $\mathcal M_2$ state-space, we will denote this representation as a $\mathcal M_{\rm 1 expanded}$ network. 
In this case it can be shown easily that $\pi_\alpha=w_\alpha/W$, where $w_\alpha$ is the weight of edge $\alpha$ and $W = \sum_\alpha w_\alpha$ \cite{Rosvall2014,Lambiotte2015}. 
The transition probability to any out-neighbour of $\alpha$ on the $\mathcal M_{\text{1 expanded}}$ network is simply
\begin{equation}
\label{mar}
T^{\mathcal M_1}_{\alpha \beta}= 
\begin{cases}
w_\beta/k_\alpha^{\rm out}      & {\rm for}~\beta \in \sigma^{\rm out}_\alpha,\cr
0      & {\rm otherwise} ,
\end{cases} 
\end{equation}
Here $\sigma^{\rm out}_\alpha$ is the set of out-neighbours of $\alpha$, and ${k_\alpha^{\rm out} = \sum_{\beta \in \sigma^{\rm out}} w_\beta}$ is the out-strength of $\alpha$.
Note that, since the original network has been undirected we have $k_\alpha^{\rm out}=k_j$, where $j$ is the endpoint of memory node (the directed edge) $\alpha=\vv{ij}$. 

\subsection*{Models of second-order Markov processes}
\label{sec:mem_models}
While second order transition matrices $T$ can be directly generated from temporal pathway data, in some cases only information about the aggregated (first-order) network might be available.
However, in such cases we can still create $\mathcal M_2$ networks using a simple memory model as described in the following, calibrated by the pathway statistics of similar datasets for which this temporal information is obtainable.
For instance, the temporal information pathways of one Email dataset may be used to fit model-parameters to generate a second-order model for different email communication network, where only aggregated information is obtainable.

\begin{figure}
\includegraphics{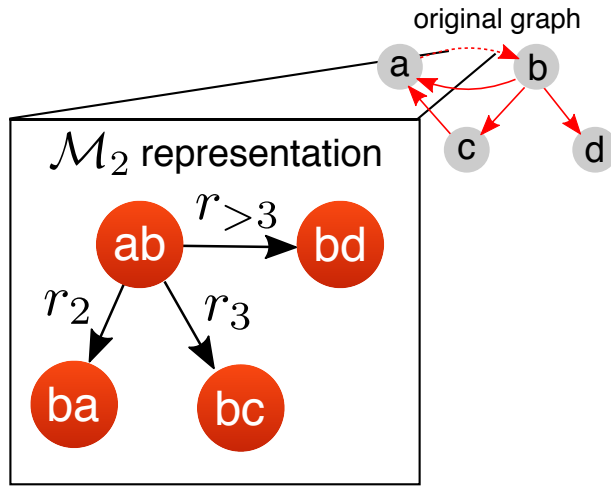}
\caption{\textbf{Construction of a $\mathcal M_2$ network using a second-order transition model.} 
    From each directed edge in the original network a walker can in principle perform three types of moves, which get weighted according to the parameters $r_i$: 
    a \textit{return step} ($r_2$), a \textit{triangular step} ($r_3$) and an  \textit{exploratory step} ($r_{>3}$). The picture shows these different options for a generic network edge.}
\label{figr}
\end{figure}

The key idea underpinning the model is to weight different types of transitions between two directed edges.
As illustrated in Figure \ref{figr}, we define three different types of transitions \cite{Rosvall2014}.
\begin{enumerate}
    \item a \textit{return step}, in which a walker coming from $\vv{ij}$ jumps to $\vv{ji}$. In other words: a walker coming from node $i$ to $j$ returns to node $i$. 
    \item a \textit{triangular step}, in which a walker coming from $\vv{ij}$ moves to edge $\vv{jk}$, where $k\neq i$ is a neighbor of $i$ .
    \item an \textit{exploratory step}, in which the walker moves from $\vv{ij}$ to an edge $\vv{jl}$, whose endpoint $l$ is neither $i$, nor any of the neighbors of $i$.
\end{enumerate}

To account for their relative importance we assign positive weights $r_2$, $r_3$ and $r_{>3}$ to the different types of transition as follows.
Let us denote the adjacency matrix of the directed line graph associated with our network by $G$.
Then we can decompose $G$ into three matrices $G^\text{ret}$ (return links), $G^\text{tri}$ (triangular), and $G^\text{exp}$ (exploratory) each containing only the links of the respective type.
To obtain a weighted memory model we now define the weighted $2M \times 2M$ adjacency matrix:
\begin{equation}
    G^{\text{mem}}_{\alpha \beta} = r_2 G^\text{ret}_{\alpha \beta} + r_3 G^\text{tri}_{\alpha \beta} + r_{>3} G^\text{exp}_{\alpha \beta},
\end{equation}
from which we can compute the associated second order transition matrix $T^{\rm model}$ simply by normalising each row to sum to $1$.

As can be easily verified, this definition implies that when we project the resulting walk onto the node space, we obtain a (first order) Markov process when $r_2=r_3=r_{>3}=\text{const}$. 
We further remark that if the dynamics is ergodic on a graph for any set of parameters, it will remain ergodic for any value of $r_2$, $r_3$ and $r_{>3}$, provided each parameter is strictly positive. 

\begin{figure*}
 \centering
 \includegraphics{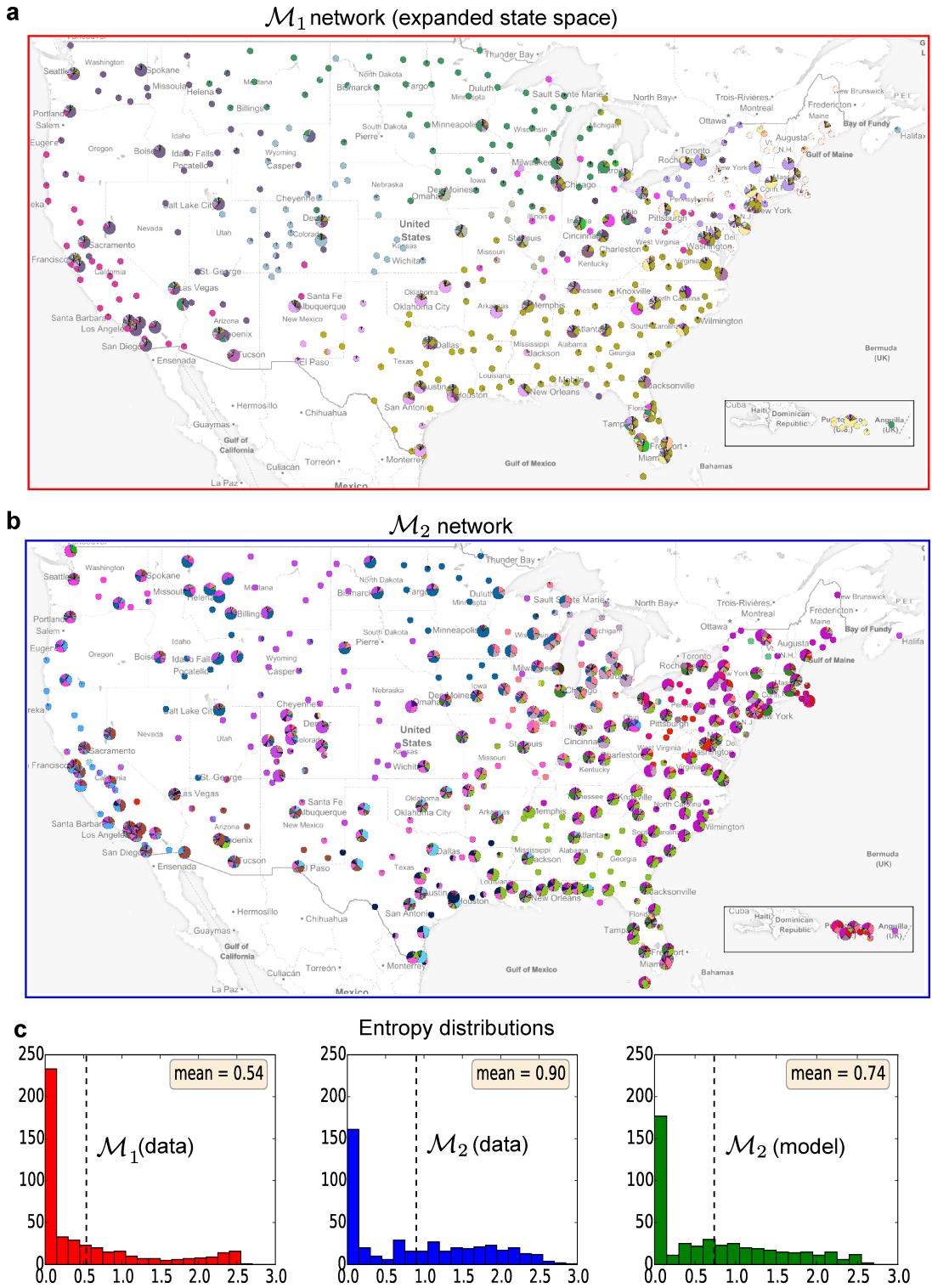}
 \caption{\textbf{Clustering analysis of a passenger traffic-network between airports in the US}. \textbf{(a)} Results based on clustering the $\mathcal{M}_{1\text{expanded}}$ system representation. \textbf{(b)} Clustering results obtained from the  $\mathcal{M}_2$ system representation. In \textbf{(a-b)} each airport is represented by a pie chart indicating the participation of the airport in different communities. For visual clarity, nodes with a community participation entropy smaller than one are displayed as smaller nodes.
     \textbf{(c)} Distributions of the community participation entropy across the network for the $\mathcal{M}_{1\text{expanded}}$ (left), $\mathcal{M}_{2 {\rm model}}$ (middle) and $\mathcal{M}_2$ (right) network. 
     All maps are created with \href{https://github.com/andrea-cuttone/geoplotlib}{Geoplotlib} using map tiles and data from \href{https://www.mapbox.com/map-feedback/}{Mapbox} and \href{http://www.openstreetmap.org/copyright}{OpenStreetMap}, \copyright OpenStreetMap contributors.}  
\label{airports}  
\end{figure*}

\subsection*{Higher order Markov dynamics reveal communities in a network of flight pathways}

Let us now demonstrate how second-order Markov dynamics can help to uncover the organisation of systems where pathway data are available, and compare their outcome with the community structure obtained by first-order dynamics and by  simplified models of second-order Markov dynamics, as defined in section \ref{sec:mem_models}.
To do so, we consider a flight network in the United States. 
The data used to construct the network consists of individual flight trajectories of people navigating between different airports in the US.
From these trajectories one can directly obtain a $\mathcal M_1$ or $\mathcal M_2$ network as discussed above and in Ref.~\cite{Rosvall2014}. 
The main purpose of our analysis here is to illustrate the differences that may arise in community detection when considering the same mobility data from different modelling perspectives.

We analysed the modular structure of this system for the following three  scenarios:
i) a first order Markov network ($\mathcal{M}_{1 \rm{expanded}}$),  viewed from the perspectives of the edges;
ii) a second order Markov network ($\mathcal{M}_2$) where the transition probabilities are directly obtained from empirical data;
iii) a second order Markov network ($\mathcal{M}_{2 {\rm model}}$), where transitions are  approximated by a simple second-order transition matrix $T^\text{model}$ (see text), whose parameters have been fitted from the data.
In each case, optimising the linearised stability for these different Markov models yields a clustering of the edges, which can be interpreted in terms of an overlapping community structure at the airport level. 
To compare the different scenarios we concentrate on the results obtained for Markov time $t=1$ (see Fig.~\ref{airports}).
Let us also note that, for each network,  only few hardly active airports do not belong to the largest SCC. For this reason, as argued in section \ref{ergodic}, we restrict the scope to nodes in this SCC.
To be more precise, we restrict the scope to the intersection of the largest SCCs of the three processes, in order to ensure that stability is well-defined for each of them.

In order to compare the communities associated to each Markov process, we first calculate the normalised variation of information  \cite{Meila2007} between the three different partitions of  $\mathcal{M}_2$ nodes obtained by optimising stability at $t=1$. 
By construction, smaller values of variation of information indicate more similar community structures. 
We obtain a variation of information of  $0.42$ between $\mathcal{M}_2$ and $\mathcal{M}_{1 \rm{expanded}}$, $0.36$ between $\mathcal{M}_2$ and $\mathcal{M}_{2 {\rm model}}$, and
$0.38$ between $\mathcal{M}_{1 \rm{expanded}}$ and $\mathcal{M}_{2 {\rm model}}$. 
These results confirm that the clusters obtained from $\mathcal{M}_{2 {\rm model}}$ provide an intermediate solution between $\mathcal{M}_2$ and $\mathcal{M}_{1 \rm{expanded}}$, with the advantage of requiring far fewer parameters than in the $\mathcal{M}_2$ case.

After having found the edge-communities in each scenario, each node can be characterised by the set of group labels of its incident edges. 
In order to quantify the apparent difference between the covers, we measure for each node the entropy of its associated group labels.
\begin{equation}
S_i = - \sum_{c} p_{i}(c) \ln p_{i}(c),
\end{equation}
where $p_{i}(c)$ is its fraction of edges assigned to community $c$. 
A value $S_i=0$ indicates that the node is associated to a single community, while higher values indicate a more diverse participation. 
The level of overlap of an edge partition can now be characterised by the distribution of entropy values on the set of nodes. 
We observe in Figure \ref{airports} that analyzing the $\mathcal{M}_2$ network results in more overlapping communities than the expanded first-order model, while the $\mathcal{M}_{2 {\rm model}}$ network is characterized by intermediate participation values. 
Similarly as observed for the map equation~\cite{Rosvall2014}, accounting for memory via second order dynamics therefore uncovers communities with a stronger overlap, in agreement with empirical observations that higher-order dynamics tends to constrain flows  within these modules.

\section*{Analyzing time-stamped temporal networks without pathway data.}

\subsection*{Second-order Markov models of time-stamped temporal networks}

Node or link activity in networked systems often exhibits non-trivial temporal patterns, such as heterogeneous inter-event times and correlations between activations times of neighbouring edges \cite{Holme2012}. 
Different network representations of such temporal data can be adopted, each associated to a different notion of temporal community. 
A first, `naive' approach is to neglect the edge-timings and work with a static network representation, where the weight of each edge corresponds to an aggregation of the activity over the whole observed time interval. 
In this case, community detection aims at grouping nodes which have the aggregated edge-weight concentrated inside modules.

Alternatively, one can represent the system by a set of (coupled) adjacency matrices $A_t$, where each $A_t$ represents the system in a short-time frame of the whole observation period. 
By applying community detection to this representation one tries to uncover meaningful structures in each time window, while enforcing some continuity between different time intervals \cite{Mucha2010}.
This approach accordingly aims at tracking the evolution of communities in the course of time.  

As we now discuss, a third notion of community is naturally associated to the representation of dynamics on temporal networks as a second-order Markov process.
In situations when the activations of neighbouring edges present temporal correlations, the dynamics of a random walker is expected to be poorly reproduced by a first-order Markov process \cite{Scholtes2014}. 
However, by finding a second-order representation of the temporal data and using the methodology introduced above, one can capture the observed temporal constraints on flow, including causal paths and a high levels of synchrony between edge activity. 

The mapping from a time series  into a second-order Markov model is realized as follows.  
We consider a system described by an ordered set of adjacency matrices $A_t$ defining the connections between nodes at time $t \in [1,T]$, where $T$ is the number of observations. 
The static representation of the temporal  network is provided by the adjacency matrix $A_{\rm static} = \sum_t A_t$. 
Each directed edge with in $A_{\rm static}$ defines a $\mathcal{M}_2$ node.
The connection strength between $\mathcal{M}_2$ nodes is now obtained by simulating pathways of a random walk process on the temporal network as follows. 
A walker is initially randomly assigned to a node. 
At every time step the walker waits until at least one edge is available for transport. 
If an edge becomes available, the walker leaves the node with probability $(1-p_{s})$ and remains on its current node with probability $p_{s}$. 
If there are multiple possible transitions, the walker takes each edge with a probability proportional to its weight. 
This process is repeated multiple times for the observed interval $[1,T]$ in order to generate sufficiently many trajectories.
From these trajectories, one can simply construct a $\mathcal{M}_2$ network by evaluating the transition frequencies between different edge pairs. 

We remark here that several other procedures could be defined to generate $\mathcal{M}_2$ networks, too. 
For instance, we could account for the duration of the intervals between time steps, or define the walker's leaving probability to be proportional to the number of available contacts.
Note that in the above construction, when $p_{s}=0$, the walker always takes the first available edge, and one recovers the dynamics studied in e.g. Ref. \cite{delvenne2015}. 
The use of non-zero values of $p_s$ introduces a source of randomness in the dynamics, which can prevent spurious effects such as an overly strong tendency for backtracking as observed in \cite{Saramaki2015}. 
However, with increasing values of $p_s$, the original ordering of events becomes less important, and the impact of the exact timings is expected to be diluted \cite{Gueuning2015}.

\subsection*{Community detection in $\mathcal M_2$ networks constructed from temporal data: an illustrative example}
\begin{figure*}
  \centering
    \includegraphics{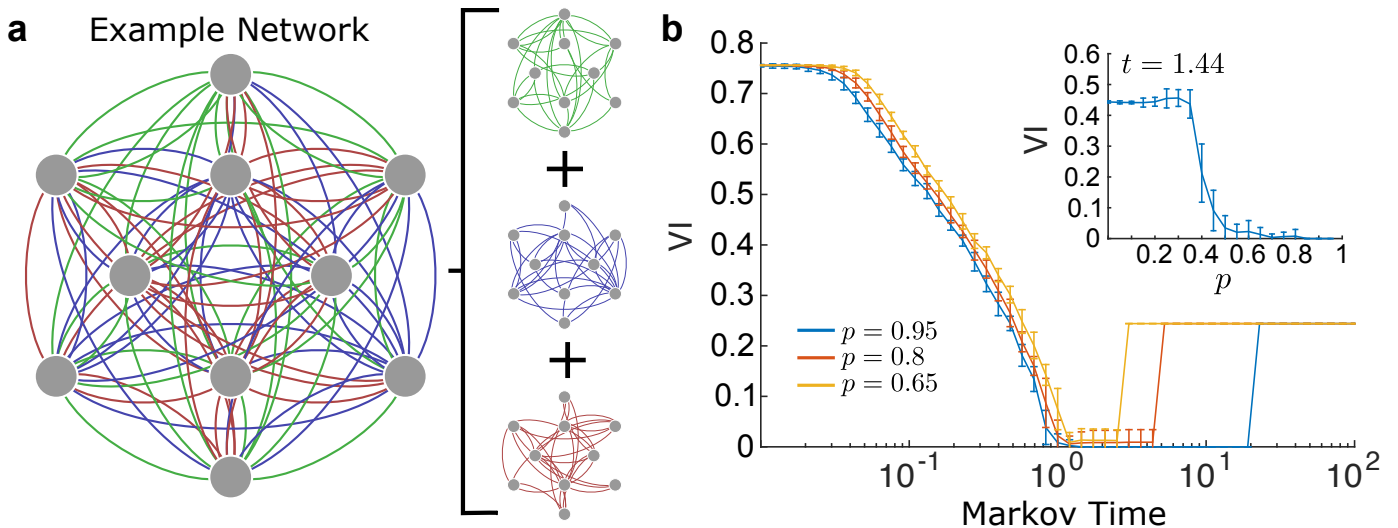}
\caption{\textbf{Analysis of an example temporal network with 10 nodes and 3 groups. }    
    \textbf{(a)} shows a complete graph with 10 nodes, whose (directed) edges are divided into 3 groups, such that each node has exactly 3 outgoing edges of each type. 
    At each time-step only edges belonging to one group are active and available for transport. 
    The currently active group ceases to be active with probability $1-p_k$, and is replaced by a randomly selected group.
    A random walker traversing the network is thus more likely to remain within the same group, the higher the probability $p_k$.
    In  panel \textbf{(b)}, the variation of information between the planted structure and the partition found by Markov stability in the $\mathcal M_2$ network is displayed as a function of Markov time. We can see that there is a plateau for values of time slightly larger than $t=1$. As shown in in the inset, the method successfully uncovers the organisation into 3 modules for sufficiently large values of $p_k$, i.e., when the groups remain active long enough in time such that the walker remains trapped inside a community for non-negligible periods.} 
    \label{ex_graph}  
\end{figure*}

Let us now show that  a $\mathcal{M}_2$ representation helps at uncovering hidden structures in temporal networks. 
As a first illustrative example, we consider a computer-generated benchmark of a social network defined as follows. 
The underlying structure is a complete graph, where the (directed) social interactions are divided into $g$ different, non-overlapping types, e.g., work relations, friends, etc. 
Fig.~\ref{ex_graph} A shows such a network with $N=10$ nodes and $g=3$ edge types.
We assume that different types of relationships dominate at different times, as often observed in empirical data \cite{Eagle2009}. 
Therefore, at each time, all edges of one single group are simultaneously active.
Initially, a randomly chosen group is active. 
At each step, the currently active group remains active with probability $p_k$; with probability $1-p_k$, a new active group is randomly selected among all groups (including the currently active one). 
For the parameters considered here, the walker thus has three outgoing edges available for transport  at each step (see Figure \ref{ex_graph}). 
It is important to note that for the generation of trajectories on this temporal network, a change of the probability for the walker to stay at its node $p_s$ is effectively equivalent to a change in $p_k$. 
For the sake of simplicity, we therefore set $p_s=0$ here.

Clearly, the trajectories of the random walker and the corresponding structure of the $\mathcal{M}_2$ network are affected by the value of $p_k$. 
When $p_k=0$, a  new active group is randomly selected at each step of the random walker. 
In this limit, the average dynamics is equivalent to a first-order random walk on a fully connected network, and thus no underlying structure can be found.
Increasing  $p_k$ implies that the random walker remains for longer time periods inside one group of edges before leaving, and it is thus possible to uncover the group structure. 
The extreme case $p_k=1$ is again trivial: the initial group remains active for all times. This case is therefore not considered in the following. 
The results of our analysis are summarised in Fig. \ref{ex_graph}, where we compare the uncovered communities with the planted solution into groups by means of normalised variation of information (VI). 
Our observations are twofold. 
First, for values of $t$ slightly larger than $t=1$, we find the planted partition with high accuracy. 
Second, one observes that the underlying communities can be successfully found when the value of $p_k$ is sufficiently large, with a clear transition around $p_k=0.4$ (see inset in Figure \ref{ex_graph}). 
 
\subsection*{A real-world case study: analyzing temporal interaction pattern of school children}
To demonstrate the utility of our approach with a real-world example, we have considered an empirical dataset from the SocioPatterns project  \cite{Gemmetto2014}. 
The data, described  in detail in  \cite{Stehle2011}, consists of time-dependent face-to-face contacts, captured by wireless wearable sensors, between children and teachers in a primary school. 
In total, the dataset is made of 77,602 contact events between 242 individuals during two consecutive days.
As the children belong to 10 different classes, we expect to find strong structural communities associated to these classes in the dynamic contact network.
In the following analysis, the pathways were generated with a probability for the walker to stay at his node at each step set to $p_s=0.05$, but similar results were obtained when varying this parameter. 

By scanning through Markov times we can find several persisting partitions.
As validation we first identified the Markov time $t_{C=10} \approx 0.631$ around which the algorithm robustly detects $10$ communities, and verified that the communities overlap with the partition into classes of the students. 
One should note here that the partition into classes can in fact be uncovered even from the aggregated, weighted network associated to the data  \cite{Stehle2011}. 

\begin{figure*}
    \includegraphics[width=\textwidth]{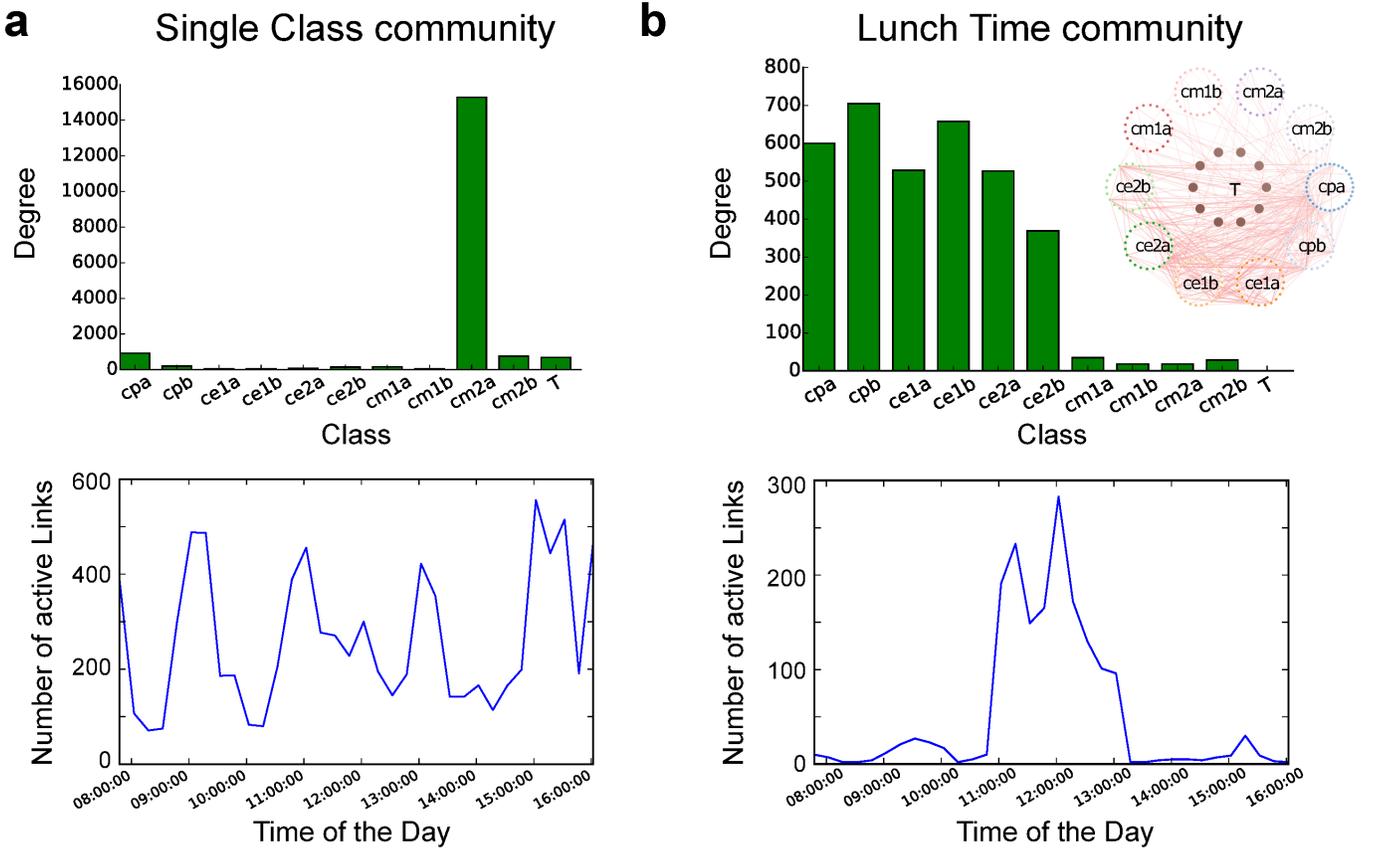}
\caption{\textbf{Analysis of the temporal community structure of a school network.} Two types of communities can be found by analyzing the $\mathcal{M}_2$ network generated with $p_{stay}=0.05$. 
Around Markov time equal to $t=0.56$ we find a robust of the directed edges into $15$ communities. $10$ of these communities are 'single class communities', in the sense that most connections are concentrated in one single class (cm2a in the example shown in \textbf{(a)}). These communities do not display a particular temporal structure in their activation patterns. This is clear from the  temporal profile of activity (bottom panel in \textbf{(a)}), which displays the number of active links as a function of time, averaged over 2 days.   
\textbf{(b)} The remaining $5$ communities are temporally localized 'lunchtime communities'. While they extend over a range of class labels, connecting pupils from different groups, they are highly coherent and synchronised in time. These communities represent the lunch-time interactions between students of different classes.}  
\label{SPstand}  \label{SPlunch}  
\end{figure*}

The advantage of our approach becomes apparent when looking at further stable partitions found for smaller Markov times. 
For instance, we find another robust split of the $\mathcal{M}_2$ network at Markov time $t=0.56$ into $15$ communities. 
For this partition most communities are still identified with classes, as can be seen in Fig.~\ref{SPstand}. 
However, there are also additional modules in which students from different classes are mixed.
These modules display activity patterns which are highly localized in time and therefore are bound to get lost when averaging over the temporal dimension in the data.
Interestingly, these latter communities correspond in fact to interactions taking place between students during lunch breaks.
Notably, these interaction are not detectable from the aggregate network or when using a memoryless representation of the dynamics (e.g., the $\mathcal M_{\rm 1 expanded}$ network).
By using the approach outlined here, we can thus not only detect 'structural' communities, in which actors maintain a higher amount of interactions between each other throughout time, but also 'temporally localized' communities, which are associated to groups that interact with each other strongly over certain time-periods.

\section*{Second-order models, non-backtracking walks and spectral clustering}

Recently, there has been an increased interest in spectral methods based on the so-called non-backtracking matrix of a network.
While standard spectral algorithms for graph partitioning tend to fail when applied to very sparse networks, Krzakala \textit{et al.} \cite{Krzakala2013} demonstrated how using the non-backtracking matrix one can design spectral algorithms which behave optimally right until the theoretical limit of detectability of a stochastic block model.
As the name suggests, the non-backtracking matrix is intimately related to the non-backtracking random walk on a network. 
The non-backtracking random walk is a diffusion process in which a particle behaves just like a simple random walker on the network, albeit with one additional constraint: after arriving at a node the walker is not allowed to immediately return to the node from which she originated (she cannot `backtrack').

Interestingly, this non-backtracking random walk can be simply phrased in terms of a memory dynamics.
Following the notation of section \ref{sec:mem_models} the non-backtracking matrix is defined on the set of $\mathcal{M}_2$-nodes and may be simply written as
\begin{equation}
    B = G^{\text{tri}} + G^{\text{exp}}.
\end{equation}
Clearly this matrix describes the line graph for a memory walker with parameters set to $r_2 = 0, r_3=1, r_{>3}=1$ (see section \ref{sec:mem_models}).
The associated transition matrix $ T^\text{B}$ of the non-backtracking random walk is  simply obtained through normalisation:
\begin{equation}
    T_{\alpha,\beta}^{B} = \dfrac{B_{\alpha \beta}}{k^{\rm out}_{\alpha}},
\end{equation}
where $k^{\rm out}_{\alpha}$ is the weighted out-degree of $\mathcal{M}_2$-node $\alpha$.
As this is just a diagonally scaled version (by the degree) of the non-backtracking operator, we may employ the flow based transition matrix $T^B$ for spectral partitioning of the nodes \cite{Newman2013a}.
In fact, $T^B$ is just one particular instance of a whole continuum of possible transition matrices, each with varying amounts of memory.
We can thus assess the effect the introduced memory has on the clustering by varying the parameters $r_i$ and performing a spectral clustering analysis.

\begin{figure}
    \includegraphics{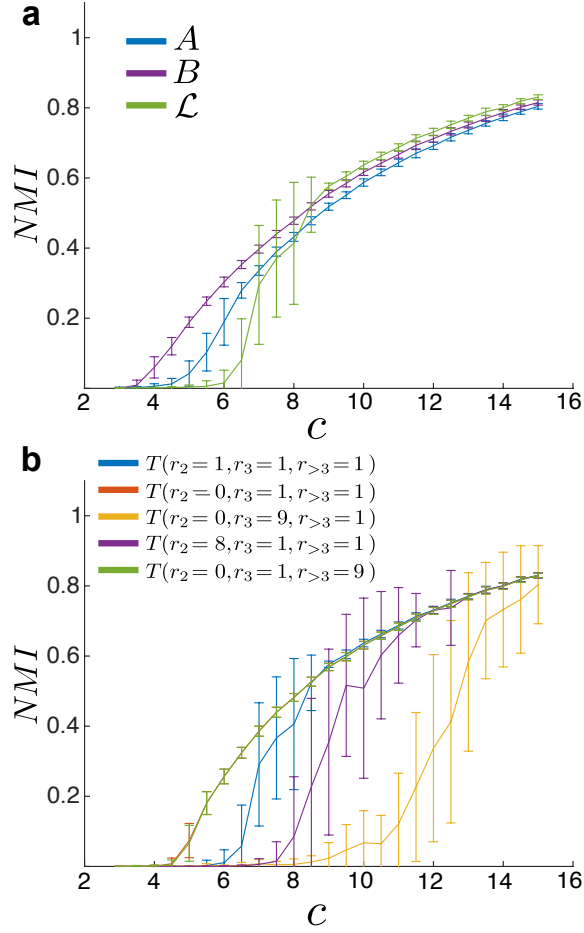}
    \caption{\textbf{Spectral Clustering of sparse networks using memory matrices.} The networks consist of $10^4$ nodes with a planted bipartition. The average degree of the network vs. the spectral clustering performance for classifying the \textit{nodes} as measured by the normalised mutual information (NMI). Note that the theoretical detectability limit is given by $c^*=\approx 3.45$ (red dashed line) \textbf{(a)} Clustering performance based on the Adjacency matrix $A$, the non-backtracking matrix $B$, and the normalised Laplacian matrix $\mathcal L$. \textbf{(b)} Clustering performance based on different $\mathcal M_{2model}$ transition matrices with different parameter settings.}
    \label{fig:spectral}
\end{figure}




Like in \cite{Krzakala2013}, we consider the problem of spectral clustering here in the simplest possible setup, which is the detection of a bipartition of equally sized groups in a large, sparse network.
The networks we consider here consist of $N=10^4$ nodes, divided into two groups according to a simple stochastic block model \cite{Krzakala2013}.
We denote the average degree of each node by $c$, the average degree to nodes inside its group by $c_{in}/2$, and the average degree to nodes outside its own group by $c_{out}/2$.
Note that these quantities are coupled by the simple relation $c=(c_{in}+c_{out})/2$.

Similar to \cite{Krzakala2013}, while keeping the ratio $c_{out}/c_{in}=0.3$ fixed, we vary the average degree $c$ and observe the results we obtain from different operators via spectral clustering.
Following \cite{Krzakala2013,Newman2013a}, each node is grouped using the following protocol.
We first compute the second dominant (left) eigenvector of the respective matrix operator. 
Second, for each we sum over the eigenvector components corresponding to all of its incoming links.
The nodes are grouped according to the sign of this sum: if is is positive the node is assigned to group 1 if not to the second group.
In case the respective matrix operator is directly defined on the node space, the nodes are simply partitioned according to the signs of the eigenvector.
Note that in contrast to the problems considered above we here look at the problem of community detection from the perspective of a hard clustering of the $\textit{nodes}$ rather than the edges and moreover assume that there is a fixed, known number of non-overlapping node-communities.
We thus do not expect that using second-order transition matrices will necessarily improve the performance (for the non-overlapping node-clustering), indeed accounting for memory can also be detrimental for this purpose as we will see in the following.
However, our main purpose here was not to design an improved node clustering algorithm but to better understand the connection between spectral clustering and memory matrices.

In Figure \ref{fig:spectral}a, we initially perform a spectral clustering analysis the non-backtracking matrix $B$ \cite{Krzakala2013}, the adjacency matrix $A$, the normalised Laplacian matrix $\mathcal L=I - D^{-1/2}AD^{-1/2}$
As has been observed by Krzakala et al., the clustering based on the backtracking matrix $B$ performs clearly better than the one based on the adjacency or Laplacian matrix, in particular when the graph becomes very sparse (see Figure \ref{fig:spectral}a).
Figure \ref{fig:spectral}b shows the results for the spectral clustering based on the memory walk matrices $T_{\{r_i\}}$ with various parameter settings $r_i$.
There appears to be a clear trend: the more weight there is assigned to exploratory steps, the easier it appears to decide on the node groupings.
Spectral clustering based on the non-backtracking second-order transition matrix $T(r_2=0,r_3=1,r_{>3=1})$ performs almost as well as the clustering based on $B$ for very sparse networks.
For networks with a higher average degree ($c\approx 8$), the results are even slightly better.
As the second eigenvector of $T$, which is used for the clustering, describes the approach of a memory dynamics towards stationarity it should not be too surprising that the results deteriorate if more memory is introduced: as very sparse graphs have a tendril like structure, additional memory can lead to a localisation of the dynamics in parts of the graph which correspond to strong 'local' bottlenecks. 
Stated differently, the slowest time-scale of the diffusion may not correlated with moving from one planted (node) community to the other, but local obstacles become more important making the second eigenvector a bad predictor of the bi-partition.

This is in particular interesting as for most memory dynamics reported \cite{Rosvall2014}, the return flow appears to be significantly large, which would imply that the dynamical constraints on the flow are much more pronounced on a local level than one may expect from the perspective of an aggregated network.
This observation appears to be aligned with the fact that many real-world systems tend to be composed of overlapping communities \cite{Ahn2010}.


\section*{Conclusions}

Network-based models have been extremely successful for the analysis of complex systems, and have led to the discovery of structural features with profound impacts on its dynamics \cite{Newman2006a}.
However, to accurately capture the complexity of real-world interacting systems, simple network models are often not sufficient.
For this reason, different approaches have  been proposed recently, which increase the model dimension and enrich the network paradigm. 
One approach that has gained prominence in the literature, is to account for the different types of interactions present in the  system by means of 
multiplex or multilayer networks \cite{Kivelae2014}. 
Our approach developed in the work presented here proceeds in a similar spirit: by providing a larger state space representation in the \textit{temporal} dimension, we aim to obtain new insights about the dynamical processes taking place in the system. 
Despite their similarities (see Ref. \cite{Lambiotte2015}), multiplex networks and higher-order Markov models differ, as the former model tends to emphasise differences in the connections of the system, while the second emphasises pathways. 

In this paper, we have focused on the applicability of second-order Markov dynamics in the context of flow-based community detection. 
Second-order Markov dynamics model real-world temporal dynamics more accurately, so the found communities are expected to better capture the actual flow constraints in the system.
In particular, we have shown how the Markov stability framework, and thus techniques like Potts models and spectral clustering, can be based on a second-order Markov dynamics, which is equivalent to considering  transitions  between directed edges in the original network.
As clustering of second order models yields a partition of the edges, one can readily interpret the results as an overlapping clustering of the nodes in the original system, which is a beneficial feature of the approach.

Some practical concerns are associated with moving to a second-order model. 
Naturally, there is an increased computational cost, as the number of nodes in the second-order model is equivalent to the number of directed edges in the original network. 
However, as most networks are sparse, this cost tends to be outweighed by the benefits gained from a higher-order dynamical representation.
Another practical issue is that pathway statistics, needed to construct a second-order transition matrix directly from data, may not always be immediately accessible.
To make the toolkit of second order models available in such scenarios, we have presented two different strategies. 
First, we have analysed a simple model able to generate realistic second-order dynamics, which can be calibrated from similar datasets.
Second, we have demonstrated how to represent temporal network data as a second order network.
We have tested these two strategies by focusing on a flight network in the United States, and a (temporal) social interaction network in a school environment, respectively.
In both cases, the second order dynamics, even approximated, allowed us to extract temporal patterns in the data that would have been missed by an aggregated first-order model.

Finally, we have highlighted the relationship between second-order Markov models and the recently introduced spectral clustering formalism based on the non-backtracking matrix.
Interestingly, the non-backtracking matrix corresponds to a scaled version of the transition matrix of a non-backtracking random walk, which is  a special case of the second order dynamics discussed in this manuscript. 
As we have demonstrated, this connection opens up the possibility to investigate further second order Markov processes and their relationships to spectral clustering.
Investigating these issues in more detail will be the subject of future work.

\section*{Acknowledgments}

We acknowledge financial support from F.R.S-FNRS, ARC and from the COST Action TD1210 KnowEscape. This paper presents research results of the Belgian Network DYSCO (Dynamical Systems, Control, and Optimismtion), funded by the Interuniversity Attraction Poles Programme, initiated by the Belgian State, Science Policy Office.

\subsection*{Author Contributions Statement}

All authors conceived the study; V.S. and M.T.S. performed the numerical simulations and created the Figures; All authors wrote and reviewed the manuscript.

\subsection*{Competing financial interests}

The authors declare no competing financial interests.


\end{document}